\documentclass[12pt]{iopart}

\usepackage{graphics,color,array,dcolumn}
\usepackage{calc}
\usepackage{amssymb}
\usepackage{xspace}

\def\gmn{{g_{\mu\nu}}}

\null

\begin{document}
\title{On the Ultraviolet Behaviour of Newton's constant}

\author{Roberto Percacci\dag, Daniele Perini\ddag}

\address{SISSA, via Beirut 4, I-34014 Trieste, Italy and
INFN, Sezione di Trieste, Italy}
\eads{\mailto{percacci@sissa.it}, \mailto{perini@he.sissa.it}}


\begin{abstract}
  We clarify a point concerning the ultraviolet behaviour of the
  Quantum Field Theory of gravity, under the assumption of the
  existence of an ultraviolet Fixed Point.  We explain why Newton's constant
  should to scale like the inverse of the square of the cutoff, even though it is
  technically inessential.  As a consequence of this behaviour, the
  existence of an UV Fixed Point would seem to imply that gravity has a
  built-in UV cutoff when described in Planck units, but not
  necessarily in other units.
\end{abstract}
\maketitle

It has often been speculated that spacetime is intrinsically grainy at
very small scales, leading to a cutoff on momentum integrations. The
most convincing arguments in favor of this idea probably come from
recent results in Loop Quantum
Gravity (for a review see \cite{Rovelli:1998yv
} and references therein), which predicts for example that the area
has a discrete spectrum with a minimum non-zero eigenvalue.  One would
like to be able to reconcile this view with the continuum, functional
methods of Quantum Field Theory (QFT) that have been so successfully
applied to other interactions.

We do not know any QFT of gravity that is both unitary and perturbatively
renormalizable, so the best hope for a consistent QFT of gravity seems to lie
in some form of nonperturbative renormalizability.
This could be realized by the existence of a nontrivial fixed point (FP)
for the Renormalization Group (RG), as suggested in
\cite{Weinberg:1979} where the notion of asymptotic safety was introduced.
Recently, some significant evidence in support of this
hypothesis has appeared in the literature
\cite{Souma:1999at,Lauscher:2001ya,
  Percacci:2002ie,
  Forgacs:2002hz}.
Let us recall briefly the features of asymptotic safety that are relevant here. 
A QFT can be defined
by an infinite set of coupling constants multiplying all possible
operators compatible with the symmetries of the theory.
In order for the theory to be considered ``fundamental'',
as opposed to an ``effective field theory'' that holds only up to a given cutoff,
it must yield finite numerical values for all physical processes at all energies.
Now, not all the couplings will be involved in the
final result of the calculation of a physical quantity.
In particular, a coupling that can be eliminated from the action by
a pointwise redefinition of the fields can also be eliminated
from the formula for a physical quantity by the same redefinition.
Such couplings are called ``inessential'' in \cite{Weinberg:1979}.
The wave function renormalization constant $Z$ is a typical example of an
inessential coupling, since it can be set to one through a constant rescaling
of the field.
If one calculates a reaction rate within
this theory and expresses it by means of such rescaled variables, one
will get a result independent of $Z$. In order for the theory to be
finite one must impose that the reaction rate be finite at all energies.
Using dimensional arguments, this implies that the
dimensionless essential couplings (that is, the essential couplings
divided by suitable powers of the energy scale, so that the ratio is
dimensionless) must go to a constant when the energy tends to infinity.
Equivalently, the corresponding beta functions have to go to zero.
Since $Z$ (as any other inessential coupling) does not appear explicitly
in the reaction rates,
no contraint needs to be imposed about its behaviour at infinite energy.

In this note we do not give any new evidence for or against the existence of a FP.
Rather, we discuss what the properties of a FP for
gravity should be \emph{if} the theory possesses it.
In particular, we clarify a point that has so far remained somewhat obscure
in discussions of asymptotic safety, namely the behaviour of Newton's constant at a FP.
Our arguments do not rely on any particular assumption about the form of
the running effective action.
The point is that, independently of the form of the action,
one can make Newton's constant disappear by a constant rescaling of the metric.
This means exactly that Newton's constant is inessential: in fact, it is
in some sense the wave function renormalization constant of the graviton.
According to the previous discussion, then, it would seem that one need
not impose any condition on the asymptotic behaviour of Newton's constant.
We discuss why, due to the peculiar character of gravity, this expectation
is incorrect and
Newton's constant must scale like the square of the inverse of the cutoff
at a gravitational FP.
Our discussion should be considered as a prequel to the work in
\cite{Souma:1999at,Lauscher:2001ya,
  Percacci:2002ie,
  Forgacs:2002hz},
where this behaviour is indeed observed in a QFT of gravity,
within specific classes of actions.

Before discussing gravity,
it is useful to review first the way in which the RG works in an
ordinary QFT in flat space.  We begin by assuming that physics at a
certain energy scale $k$ can be accurately described by a local
effective action $\Gamma_k$ containing a finite number of terms, with
coupling constants $g_i$.  One is interested in calculating the
effective action at some lower energy scale $k/a$ with $a>1$.  This
can be done by performing a functional integration over fluctuations
of the fields with momenta $q$ in the range $k/a<|q|<k$\ \footnote{We
  consider a Euclidean theory.}, using $\Gamma_k$ as the bare action
in the path integral.  In general, this functional integral will
produce infinitely many effective couplings, but under suitable
conditions the result can again be well approximated by an action of
the same form, with the only difference that the coupling constants
$g_i$ have different values.  Taking the limit $a\to 1$ one can
compute the beta functions $\beta_i=\partial_t g_i$, where
$t=\log(k/k_0)$, for some arbitrary $k_0$.  These beta functions
depend in general on all couplings $g_j$.  Integrating the
corresponding first order differential equations produces
one-parameter families of effective actions $\Gamma_k$, all of the
same form but with running coupling constants.  In some cases the
theory breaks down at some scale; it is then to be regarded just as an
effective field theory. In others, it is possible to take the limit
$k\to\infty$; one can then regard the theory as being ``fundamental''.

There are two technical points to be taken into account at this stage.
First, dimensionful quantities do not have a measurable value: one can only
measure their ratio to some other quantity of the same dimension.
Therefore, in the RG, it is customary to take the cutoff $k$ as a unit
of mass and to consider the beta functions of the dimensionless
ratios $\tilde g_i=g_i/k^{d_i}$, where $d_i$ is the mass dimension
of the coupling $g_i$.

Second, as mentioned earlier, one can eliminate the inessential couplings
from the action by means of field redefinitions.
For example in the case of the wave-function renormalization constant $Z$
one can always rescale the field (and the remaining couplings) in such
a way that $Z=1$. The functional integral over an infinitesimal
momentum shell however modifies the value of $Z$, so if we want to
maintain $Z=1$ at all scales, it is necessary to rescale the fields at every
RG step.

An infinitesimal RG transformation therefore consists of the
following three steps \cite{Wilson:1983dy}: 1) an integration over
an infinitesimal shell of momenta from $k$ to $k-dk$ ({\it i.e.}
$a=1+\delta t$); 2) a rescaling of all momenta by a factor
$1-\delta t$ so that $k$ is restored as a unit of momentum; 3) a
redefinition of the fields to restore the values of the
inessential couplings.

Since we are interested in applications of this formalism in a
gravitational context, it is more convenient to view the second step
as a rescaling of the metric. This is equivalent to a rescaling of
coordinates and momenta in the case of a flat metric, but generalizes
also to curved metrics. Let us illustrate this point in the specific
case of a single scalar field $\phi$ with an action of the form
$$
S(\phi,\gmn;Z_\phi,m^2,\lambda,\ldots)=
\int d^4x\sqrt{g} \left[
{Z_\phi\over 2}g^{\mu\nu}\partial_\mu\phi\partial_\nu\phi
+{1\over 2}m^2\phi^2+\lambda\phi^4+\ldots
\right]\ ,\eqno(1)
$$
where the ellipses stand for other terms with higher powers of the
field and/or higher derivatives.
(We are not interested here in the issue whether this theory
actually has a FP, but only in the methodology.)  The metric plays the
role of an external field and can be assumed to be flat.  We assume
the coordinates to be dimensionless and $\gmn$ to have dimension of
length squared.  This action is invariant under the following
two-parameter family of rescalings:
$$
S(\phi,\gmn;Z_\phi,m^2,\lambda,\ldots)=
S(bc\phi,b^{-2}\gmn; c^{-2} Z_\phi, b^2 c^{-2}m^2, c^{-4}\lambda,\ldots)\ .\eqno(2)
$$
The power of $c$ in front of a coupling counts the power of $\phi$
appearing in the corresponding term of the action.
The power of $b$ in front of a coupling counts twice the power
of $\gmn$ minus the power of $\phi$ appearing in the
corresponding term in the action.
Note also that the power of $b$ in front of any quantity
is equal to its canonical dimension.
In fact this {\it is} the definition of the
canonical dimension of every quantity appearing in eq.\ (1).

If one scales all quantities, including the cutoff, in the appropriate
way, invariance under scalings can be extended to the quantum theory.
Assuming that it can be approximated by an action of the form (1), the
running effective action $\Gamma_k$ can be shown to have the
invariance
$$
\Gamma_k(\phi,\gmn;Z_\phi,m^2,\lambda,\ldots)=
\Gamma_{bk}(bc\phi,b^{-2}\gmn;c^{-2} Z_\phi, b^2 c^{-2}m^2, c^{-4}\lambda,\ldots)\ ,\eqno(3)
$$
with $b$, $c$ real numbers.  This can be generalized to arbitrarily
complicated actions.  Note that the cutoff behaves just like another
dimensionful coupling.  One can use the $c$-invariance to fix for
example $Z_\phi=1$ and the $b$-invariance to fix $k=1$. One can then
define
$$
\tilde\Gamma(\tilde\phi,\tilde g_{\mu\nu},\tilde m^2,\tilde\lambda,\ldots)=
\Gamma_1(\tilde\phi,\tilde g_{\mu\nu} ,1, \tilde m^2, \tilde\lambda,\ldots)
=\Gamma_{k}(\phi,\gmn ,Z_\phi,  m^2, \lambda,\ldots)\ ,\eqno(4)
$$
where $\tilde\phi=\sqrt{Z_\phi}\phi/k$, $\tilde m^2=m^2/(k^2 Z_\phi)$, $\tilde\lambda=\lambda/Z_\phi^2$,
$\tilde g_{\mu\nu}=k^{2}\gmn$.

With these definitions, the second step in a RG transformation is a
scaling with a parameter $b=1-\delta t$, which brings $k$ back to its
original value, and the third step is a scaling with a parameter
$c=1-{1\over 2}\delta t\eta_\phi$, with $\eta_\phi={\partial_t
  Z_\phi\over Z_\phi}$ which brings $Z_\phi$ back to its original
value.  These transformations produce a flow for the action
$\tilde\Gamma$.

When one computes the beta functions of the essential couplings, they
can be written as functions of the essential couplings and of the
anomalous dimension $\eta_\phi$.  There is no explicit dependence on
$Z_\phi$ or $k$.  A FP for the scalar field would be a point where the
beta functions of the essential couplings vanish.  The beta function
of $Z_\phi$ is not required to vanish; instead, the anomalous
dimension $\eta_\phi$ at the FP can be computed once the values of the
essential couplings at the FP are known.

Up to a point, things work in the same way also for gravity. There are
no obstacles to performing quantum calculations in gravity as long as
one regards it as an effective field theory
\cite{Donoghue:1994dn
}. One can also write exact, nonperturbative RG equations for gravity:
since the range of momenta to be integrated over in a RG step is
finite, there are no divergences in the definition of the beta
functions. By means of suitable approximations, one can thus obtain
equations that describe the evolution of the gravitational couplings
for finite $k$ \cite{Reuter:1998cp,Dou:1998fg}. The issue is
whether it is possible to consistently let $k$ go to infinity.

For the sake of definiteness consider pure gravity with the following action:
$$
S(\gmn;\Lambda,Z_g,\ldots)=
\int d^4x\sqrt{g} \left[2Z_g\Lambda-Z_gR[g]+\ldots\right]\ ,
\eqno(5)
$$
where $Z_g=(16\pi G)^{-1}$, $\Lambda$ is the dimension-two
cosmological constant and the dots stand for higher powers of curvature.
As discussed earlier, one should treat
separately the essential and inessential gravitational couplings. The
coupling $Z_g$ has the same role as the wave function renormalization
$Z_\phi$.  In the linearized theory, it multiplies the kinetic term of
the graviton, and it can be entirely eliminated from the action by a
rescaling of the metric\footnote{In the linearized theory, the
  background and the fluctuation have to be rescaled in the same way.}
and therefore is an inessential coupling\footnote{It
  has been argued that perhaps Newton's constant is essential because
  of boundary terms in the gravitational action
  \cite{Gastmans:1978ad}. One can avoid this type of arguments
  assuming that the fields fall off sufficiently fast at infinity.}.

The action (5) has the following scaling property:
$$
S(\gmn;\Lambda,Z_g,\ldots)=
S(b^{-2}\gmn;b^2\Lambda,b^2 Z_g,\ldots)\ .
\eqno(6)
$$
As with scalar fields, this turns into a scaling property of the
running effective action:
$$
\Gamma_k(\gmn;\Lambda,Z_g,\ldots)=
\Gamma_{bk}(b^{-2}\gmn;b^2\Lambda,b^2 Z_g,\ldots)\ .
\eqno(7)
$$

Following the example of the scalar field, we can use this freedom
to set $k=1$. We define
$$
\tilde\Gamma(\tilde g_{\mu\nu};\tilde Z_g,\tilde\Lambda,\ldots)=
\Gamma_1(\tilde g_{\mu\nu};\tilde Z_g,\tilde\Lambda,\ldots)=
\Gamma_k(\gmn;Z_g,\Lambda,\ldots)\ ,
\eqno(8)
$$
where
$$
\tilde g_{\mu\nu}=k^{2}\gmn\ ;\ \
\tilde Z_g=\frac{Z_g}{k^2}=\frac{1}{16\pi \tilde G}\ ;\ \
\tilde\Lambda=\frac{\Lambda}{k^2}\ .
\eqno(9)
$$
If this were any other QFT, it would also be possible to
eliminate the inessential coupling $Z_g$.
But here the peculiar character of gravity emerges.
In the case of the scalar field (or any other field theory), the
scaling of the field (parameter $c$ in eq.\ (3)) and the scaling of the
momenta (parameter $b$ in eq.\ (3)) are two independent operations.  In
the case of gravity, since the length of any vector (in particular the
modulus of the momentum) is determined by $\gmn$, the scaling of the
field and the scaling of the momenta are the same operation.
This is why in eqs.\ (6--7) we have only a one-parameter scaling, while in (2--3) we
had a two-parameter scaling.
As a result of this fact, in the case of gravity it is impossible to eliminate
at the same time $k$ and $Z_g$ from the action.

If we choose to eliminate $k$ as in (8), the effective action and
also the beta functions of all couplings will depend on $Z_g$ (or equivalently
on Newton's constant $G$).
As a consequence, at a FP, one has to require the beta function of $\tilde G$ to vanish,
along with those of all the essential couplings.

There is also an alternative procedure:
one can use the $b$-freedom to set $Z_g=Z'_g$ for
some arbitrary fixed value $Z'_g$.
Then we can define a new action depending only
on $\Lambda'$ and the other essential couplings:
$$
\Gamma'_{k'}(g'_{\mu\nu};\Lambda',\ldots)=
\Gamma_{k'}(g'_{\mu\nu};\Lambda',Z'_g,\ldots)=
\Gamma_k(\gmn;\Lambda,Z_g,\ldots)\ ,
\eqno(10)
$$
where
$$
g'_{\mu\nu}={Z_g\over Z'_g}\gmn\ ;\
\Lambda'={Z'_g\over Z_g}\Lambda\ ;\ k'=\sqrt{Z'_g\over Z_g}k\ . \eqno(11)
$$
There does not seem to be any reason to assume that $b$ is dimensionless,
therefore one can choose $Z'_g=1/16\pi$,
which amounts to working in Planck units.

If we choose to eliminate $Z_g$ as in equation (10),
the beta functions for the essential couplings will retain an
explicit dependence on the variable $k'$.  Reexpressing the
derivatives in terms of the independent variable $t'=\log k'$:
$$
\partial_{t'}\Lambda'={2\over 2-\eta}\partial_t\Lambda'=
\beta_{\Lambda'}(\Lambda',k',\ldots),\eqno(12)
$$
where $\eta=\partial_t Z_g/Z_g$ and the ellipses stand for
all the essential couplings of the theory.
This behaviour can indeed be verified for the beta functions given for example in
\cite{Reuter:1998cp,Dou:1998fg}, and is most clear in those cases when
the cutoff is chosen in such a way that the momentum integrals can be
performed explicitly, as for example in \cite{Litim:2003vp}.

Thus, if we choose to work in Planck units we obtain a non-autonomous system of
RG flow equations. This is unlike any other field theory, where
the beta functions never depend explicitly on the cutoff.
The explicit $k'$-dependence makes the very notion of FP a priori somewhat unclear,
since the evolution of the couplings has to follow a time-dependent
vectorfield: a point where the beta function vanishes instantaneosly
is not a FP in general because the zero moves.

In practice, the simplest way to solve for the flow defined by eq.\ (12)
is to exploit the relations between the variables and map the flow of $\tilde\Lambda$ and
$\tilde G$ as a function of $t$ onto the flow of $\Lambda'$ as a
function of $t'$.
Referring now to the explicit calculations in
\cite{Souma:1999at,Lauscher:2001ya,
Percacci:2002ie,
Forgacs:2002hz,Reuter:2002kd}, when one actually solves for the flow,
the picture is slightly more complicated due to the fact that the flow
of $\tilde\Lambda$ and $\tilde G$ near the FP follows a spiralling
motion. Thus, $k'$ does not grow monotonically with $k$: it first
overshoots its FP-value $k'_*$ and then approaches it with damped oscillations. As
a consequence, $k'$ is not a good independent variable for describing
the flow of $\Lambda'$ (it can only be used on finite intervals where
${dk'\over dk}\not= 0$).

This problem can be circumvented by using a different independent variable,
that can be constructed as follows.  In a suitable coordinate system
in a neighborhood of the FP in the $\tilde\Lambda$-$\tilde G$ plane,
the linearized vectorfield is given by $ \partial_t
g_i(t)=M_{ij}g_j(t), $ where $M_{11}=M_{22}=-\alpha$ and
$M_{21}=-M_{12}=\omega$, with $\alpha\approx 1.86$ and $\omega\approx
4.08$. There is no real linear transformation that diagonalizes $M$.
In polar coordinates $\rho$, $\varphi$ the solutions of this equation
are $\rho(t)=\rho_0\, e^{-\alpha t}$, $\varphi(t)=\varphi_0+\omega t$.

One can perform a diffeomorphism of the $\tilde\Lambda$-$\tilde G$
plane which undoes the spiralling motion, such that the new $\tilde G$
variable is a monotonic function of $t$.  We cannot write the explicit
form of this diffeomorphism, but in the neighborhood of the FP it is
$\rho'=\rho$ and $\varphi'=\varphi+{\omega\over\alpha}\log(\rho)$.
The transformation is singular at the FP and can be joined smoothly to
the identity far from the FP. Denoting with $y$ and $x$ the Cartesian
coordinates corresponding to the polar coordinates $\rho'$ and
$\varphi'$, the flow equations become simply $\frac{dx}{dt}=-\alpha x$
and $\frac{dy}{dt}=-\alpha y$.  For $t\to -\infty$, $x\to \tilde G$
and $y\to\tilde\Lambda$.  Except for the line $x=0$, one can now take
$x$ as the new independent variable, in which case eq.\ (10) takes the
simple form $\frac{dy}{dx}=\frac{y}{x}$.  The trajectories of the
class Ia and IIa in the terminology of \cite{Reuter:2002kd} lie in the
region $x<0$ and have ${dx\over dt}>0$, while those of the class IVa
lie in the region $x>0$ and have ${dx\over dt}<0$.

Let us now return to the more familiar picture in cutoff units.
The fact that Newton's constant behaves in some sense like an essential coupling,
has the following consequences.
On one hand, in spite of not
being an essential parameter, $\tilde G$ has to reach a finite limit
$\tilde G_*$ at an UV FP. Now we observe that $k'=\sqrt{\tilde G}$,
{\it i.e.}  the cutoff in Planck units is (the square root of)
Newton's constant in cutoff units.  Consequently, if gravity had a FP,
the cutoff would have a finite limit
$$
k'_*=\sqrt{\tilde G_*}\eqno(13)
$$
in Planck units.

On the other hand, since
$$
\eta=\frac{\partial_t
  \tilde{Z}_g}{\tilde{Z}_g}+2 \ , \eqno(14)
$$
the anomalous dimension $\eta$ has to be exactly equal to 2 at a
non-Gaussian FP \cite{Lauscher:2001ya}, suggesting that at very small
scales the world may look two-dimensional.

When matter is added to the picture additional possibilities arise,
giving rise to interesting scenarios
\cite{Percacci}.
We found in
\cite{Percacci:2002ie
}
that adding a scalar field with interactions of the form $\phi^{2n}$
and $\phi^{2n}R$ does not alter significantly the result of pure
gravity, in the following sense: under the same approximations
that produce a FP for pure gravity, there is still a FP, with slightly
shifted values of $\Lambda_*$ and $G_*$ and all scalar couplings
with $n\geq 1$ equal to zero.

Now, ordinary systems of units are based on atomic spectroscopy.
The scale of the energy levels of atoms is set by the mass of the electron,
which is in turn dictated by the VEV of the Higgs field.
In this sense, ordinary units are based on the mass of the Higgs field.
If the system admits a FP, the results of
\cite{Percacci:2002ie
}
suggest that the ratio $\upsilon^2/k^2$ goes to zero at this FP,
$\upsilon^2$ being the VEV of the scalar field.
Therefore, if we interpret $\phi$ as the modulus of the Higgs field,
there would be no minimal length in Higgs units.
On the other hand (13) can be taken as an indication that the theory has a minimal
length in Planck units.
These remarks show that the answer to the question whether gravity
has a minimal length could depend upon the system of units.

\ack
We would like to thank M.\ Reuter for correspondence on the
essentiality of Newton's constant.

\end{document}